\documentstyle[12pt,thmsa,sw20lart]{article}

\input{tcilatex}
\begin{document}

\author{C. Bagnuls\thanks{%
Service de Physique de l'Etat Condens\'{e}} \ and C.\ Bervillier\thanks{%
Service de Physique Th\'{e}orique e-mail: bervil@spht.saclay.cea.fr} \\
C. E. Saclay, F91191 Gif-sur-Yvette Cedex, France}
\title{Nonasymptotic critical behavior from field theory}
\date{December 14, 2000}
\maketitle

\begin{abstract}
The obtention (up to five or six loop orders) of nonasymptotic critical
behavior, above and below Tc, from the field theoretical framework is
presented and discussed.
\end{abstract}

When talking about critical behavior one usually thinks of critical
exponents (power laws), and eventually of corrections to scaling, all
notions strictly related to the unprecise definition of an asymptotic
critical domain. In fact criticality may be observed beyond that theoretical
domain and, sometimes, this makes it difficult to compare theory and
experiments.~\cite{223} For example, it is thought that some systems could
undergo a retarded crossover~\cite{4003} from classical to Ising-like
critical behaviors. In such a case, the critical domain would be much larger
than for, say, pure fluids. Consequently many correction-to-scaling terms
should be introduced and, it is very likely that the series would not
converge. For that reason, nonasymptotic theoretical expressions of critical
behaviors are required to describe such systems.

It is not very well known that, beyond the estimations of the critical
exponents, the renormalization group (RG) theory~\cite{440} is also adapted
to provide us with nonasymptotic forms of the critical behavior especially
when a crossover phenomenon occurs (the crossover is then characterized by
the competition of two fixed points).

We briefly present here the principles of the calculations done within the
massive field theoretical framework in three dimensions ($d=3$)~\cite{340}
and which have yielded accurate nonasymptotic forms of the susceptibility $%
\chi (\tau )$ and the specific heat $C(\tau )$ for $\tau =(T-T_{{c}})/T_{{c}%
}>0$ and $\tau <0$, of the correlation length $\xi (\tau )$ for $\tau >0$
and of the coexistence curve $M(\tau )$ for $\tau <0$.~\cite{733,736} The
calculations presented here have induced, directly or indirectly, several
subsequent works.~\cite{754,3639} We hope that this text will encourage
further works on nonasymptotic critical behavior. We think, in particular,
that the variational perturbation theory used recently to estimate universal
exponents~\cite{3998} and amplitude ratios,~\cite{4937} could be an
advantageous tool.

Let us first specify the meaning of the title. ``Nonasymptotic critical
behavior'' means that we perform a resummation of the infinite series of
correction-to-scaling terms which are expected~\cite{413} in the asymptotic
expression of any singular quantity such as $\xi (\tau ).$ Particularly, for 
$\tau \rightarrow 0^{+,-}$, we have: 
\begin{equation}
\xi (\tau ) =\xi _{0}^{+,-}\left| \tau \right| ^{-\nu }\left[
1+\sum_{n=1}^{\infty }\sum_{m=1}^{\infty }a_{\xi ^{+,-}}^{\left( n,m\right)
}\left| \tau \right| ^{n\Delta _{m}}\right]  \label{XiTot}
\end{equation}
in which $\nu $ is a critical exponent, $\xi _{0}^{+,-}$ stands for the
leading critical amplitudes in the two phases and the coefficients $a_{\xi
^{+,-}}^{\left( n,m\right) }$ correspond to the amplitudes of the confluent
corrections to scaling controlled by the exponents $\Delta _{m}$ ($%
m=1,2,\cdots ,\infty $). Those exponents ($\nu $ and $\Delta _{m}$) arise in
a linear study~\cite{413} of the RG transformation in the vicinity of a
fixed point: the solutions of the eigenvalue problem provide us with some
positive (say one, $\lambda _{0}$, for simplicity\footnote{%
The number of positive eigenvalues depends on the fixed point considered.
Since we are interested in a critical point (we mean not a multi-critical
point) then there is only one positive eigenvalue and the fixed point
referred to in that case is the famous Wilson-Fisher~\cite{439} fixed point.}%
) and infinitely many negative eigenvalues ($\lambda _{m}$ for $m=1,2,\cdots
,\infty $). Then we have $\nu =1/\lambda _{0}$, $\Delta _{m}=-\nu \lambda
_{m}$ ($m=1,2,\cdots ,\infty $). The case $m=1$ corresponds to the first
correction-to-scaling term associated to the largest negative eigenvalue and
the usual notations are $\omega $ (for $-\lambda _{1}$) and $\Delta =\omega
\nu $ (for $\Delta _{1}$). In a linear study of the RG, discarding the
next-to-leading correction terms in Eq. (\ref{XiTot}),\ it is usual to
write: 
\begin{equation}
\xi (\tau )\simeq \xi _{0}^{+,-}\left| \tau \right| ^{-\nu }\left[ 1+a_{\xi
}^{+,-}\left| \tau \right| ^{\Delta }\right]  \label{XiAsymp}
\end{equation}
with the universality of the ratios $\xi _{0}^{+}/\xi _{0}^{-}$ and $a_{\xi
}^{+}/a_{\xi }^{-}$. Eq. (\ref{XiAsymp}) is only valid asymptotically close
to the critical point.

Notice that the infinite sum in Eq. (\ref{XiTot}) does not converge for
large values of $\tau $. Thus, to get a useful nonasymptotic expression of
the critical part of $\xi $, we must consider a resummation procedure. It is
provided to us by the RG theory. However the framework we use implies an
approximation: ``from field theory'' means that only one family of
correction-to-scaling terms (associated to $m=1$) is accounted for. Thus,
instead of Eq. (\ref{XiTot}), our effective expression for $\xi (\tau )$ is: 
\begin{equation}
\xi _{FT}(\tau )=\xi _{0}^{+,-}\left| \tau \right| ^{-\nu }\left[
1+\sum_{n=1}^{\infty }a_{\xi ^{+,-}}^{\left( n\right) }\left| \tau \right|
^{n\Delta }\right]  \label{XiFT}
\end{equation}
As one could see from the use of our calculations in analyzing experimental~%
\cite{734} or Monte-Carlo~\cite{4732} data, the approximation of field
theory does not prevent the study from yielding physically useful
nonasymptotic critical behaviors. In fact, in the case of Eq. (\ref{XiFT}),
the range $0< \tau < \infty $ corresponds to an {\em interpolation between
two fixed points} and, consequently, the crossover is described by universal
functions.~\cite{4073}

\section{Principles of the calculations}

One starts from the ``bare'' or unrenormalized $\phi ^{4}$-hamiltonian in $d$
euclidean dimensions corresponding to the scalar field theory to be
renormalized: 
\begin{equation}
H=\int {d}^{d}x\;\left[ \frac{1}{2}\left\{ \left( \nabla \phi _{0}\right)
^{2}+r_{0}\phi _{0}^{2}\right\} +\frac{g_{0}}{4!}\phi _{0}^{4}\right]
\label{Hamil}
\end{equation}
in which it is implicit that $\phi _{0}$ depends on $x$ and may eventually
represents a vector with $n$ components in which case Eq. (\ref{Hamil}) is
supposed to satisfy the $O(n)$-symmetry. The bare coupling $g_{0}$ is
dimensionful and ``measured'' in unit of $\Lambda $ (the ultra-violet
cutoff): 
\[
g_{0}=u_{0}\Lambda ^{\epsilon } 
\]
in which $\epsilon =4-d$ and $u_{0}$ is dimensionless.

There exists two kinds of renormalization scheme for the scalar field
theory: the massive and the Weinberg~\cite{2272} schemes. In the massive
scheme, the unit of reference is provided by the mass parameter $m=\xi ^{-1}$%
. In this framework the critical theory, corresponding to $m=0$, is not
defined. On the contrary, in the Weinberg scheme, one first defines the
critical theory (massless theory) and the unit of length scale is provided
by the inverse of some arbitrary momentum-subtraction-point parameter $\mu $%
. In that renormalization scheme, the ``soft-mass'' parameter $t$ is
introduced via the renormalization of insertions of the $\phi _{0}^{2}$%
-operator within the vertex-functions; when it is different from zero, $t$
is (linearly) proportional to the reduced temperature scale $\tau $ defined
above. Though we have effectuated our calculations within the massive
framework --- because the longest available\footnote{%
The unpublished Guelph report~\cite{323} may be obtained via H. Kleinert and
V. Schulte-Frohlinde web site at http://www.physik.fu-berlin.de/\-\symbol{126%
}kleinert/kleiner\_reb8/\-programs/programs.html.} perturbative series~\cite
{323} have been obtained within the massive scheme directly in $d=3$ ---,
the presentation of the principles of the calculation is simpler within the
Weinberg scheme. In that scheme, the renormalization conditions correspond
to the following (re)-definitions: 
\begin{eqnarray}
\phi _{0} &=&\left[ Z_{3}\left( u\right) \right] ^{1/2}\phi  \label{W1} \\
\left( \phi _{0}\right) ^{2} &=&\frac{Z_{3}\left( u\right) }{Z_{2}\left(
u\right) }\left( \phi \right) ^{2}  \label{W2} \\
u_{0}\Lambda ^{\epsilon } &=&\mu ^{\epsilon }u\frac{Z_{1}\left( u\right) }{%
\left[ Z_{3}\left( u\right) \right] ^{2}}  \label{W3} \\
r_{0} &=&r_{0{c}}+\frac{Z_{2}\left( u\right) }{Z_{3}\left( u\right) }t
\label{W4}
\end{eqnarray}
in which $u$ is the renormalized coupling and $r_{0{c}}$ is defined by:

\begin{equation}
\left. \Gamma _{0}^{(0,2)}\left( p;r_{0},g_{0}\right) \right|
_{p=0,r_{0}=r_{0{c}}}=0  \label{R0c}
\end{equation}
in which the subscript 0 refers to the bare theory.

{\em Up to analytical terms} which are usually neglected when studying
critical phenomena, the quantity $r_{0}-r_{0c}$ is proportional to the
physical parameter $\tau $: 
\begin{equation}
\frac{r_{0}-r_{0{c}}}{\Lambda ^{2}}=\theta \tau +O\left( \tau ^{2}\right)
\label{Atau}
\end{equation}
in which $\theta $ is a nonuniversal factor.

The renormalized $N$-point vertex-functions with $L$-insertions of the $%
\left( \phi \right) ^{2}$ operator are related to their unrenormalized
counter parts as follows:

\[
\Gamma ^{(L,N)}\left( \left\{ q,p\right\} ;u,\mu \right) =\left[ Z_{3}\left(
u\right) \right] ^{N/2}\left[ \frac{Z_{2}\left( u\right) }{Z_{3}\left(
u\right) }\right] ^{L}\Gamma _{0}^{(L,N)}\left( \left\{ q,p\right\}
;r_{0},g_{0}\right) 
\]

That the renormalization functions $Z_{i}\left( u\right) $ be defined by
renormalization conditions on the 2-point and 4-point vertex-functions
considered at some subtraction momentum point expressed in terms of the only
dimensionful (momentum-like) parameter $\mu $ or by a ``minimal''
subtraction procedure (such as the subtraction of poles located at $\epsilon
=0$) does not matter for the following.

In field theory, the RG originates from the arbitrariness of the subtraction
procedure for a given bare theory. Hence, the renormalized quantities $u$
and $t$ become functions of the renormalization parameter $l=-\ln \left( \mu
/\Lambda \right) .$ Consequently, Eq. (\ref{W4}) must be understood as
follows: 
\begin{equation}
r_{0}=r_{0{c}}+\frac{Z_{2}\left[ u\left( l\right) \right] }{Z_{3}\left[
u\left( l\right) \right] }t\left( l\right)  \label{W4bis}
\end{equation}

Now, by imposing that $t\left( l\right) $ remains a fixed quantity (say $%
t\left( l\right) =1$), one relates the evolution of $u\left( l\right) $ to
the approach to the critical point of the bare (physical) theory (defined by 
$r_{0}\rightarrow r_{0{c}}$). Then, with $t\left( l\right) =1$, Eq. (\ref
{W4bis}) shows that, for $r_{0}=r_{0{c}}$, $u\left( l\right) $ must take on
a particular value $u^{*}$ so that $Z_{2}\left( u^{*}\right) /Z_{3}\left(
u^{*}\right) $ vanishes. Of course, $u^{*}$ is the nontrivial zero of the
famous $\beta $-function: 
\begin{equation}
\beta \left( u\right) =-\left. \frac{{d}u\left( l\right) }{{d}l}\right|
_{u_{0}}  \label{beta}
\end{equation}
with a positive $\left. \omega =d\beta \left( u\right) /du\right| _{u=u^{*}}$
so that $u\left( l\right) \stackrel{l\rightarrow \infty }{\rightarrow }u^{*}$%
.

The pure scaling (power law) regime of vertex-functions corresponds to $%
u\left( l\right) =u^{*}$ and the first correction-to-scaling term [as in Eq.
(\ref{XiAsymp})] to a linear correction proportional to $u\left( l\right)
-u^{*}$. As $u\left( l\right) $ moves further away from $u^{*}$, more and
more correction terms must be included but then a nonlinear study is
required. It is a matter of fact that the domain $0<u\left( l\right) <u^{*}$
corresponds to the entire domain $\infty >r_{0}-r_{0{c}}>0$. Therefore, if
one re-sums perturbative series in powers of $u\left( l\right) $ in the
range $\left] 0,u^{*}\right[ $, one implicitly obtains nonasymptotic
critical answers which interpolate \footnote{%
We have also performed calculations~\cite{3458} for $u>u^{*}$. In this
range, the sign of the first correction-to-scaling term is changed and
corresponds to the Ising model~\cite{38}.} between a classical critical
behavior (when $u\left( l\right) $ is small) and, say, an Ising-like
critical behavior (for the $O(1)$-symmetry) when $u\left( l\right) $
approaches $u^{*}$. It remains to invert Eq. (\ref{W4bis}) to express these
answers under the forms of functions of $r_{0}-r_{0{c}}$ [or of $\tau $, via
Eq. (\ref{Atau})] which is the genuine physical ``measure'' of the distance
to the critical point.

With a view to get the best possible accuracy, we have looked at the
available calculations up to relatively high orders of perturbation series.\
There are two kinds of such calculations:

\begin{enumerate}
\item  analytically up to fifth order in the Weinberg scheme with
dimensional regularization and minimal subtractions.~\cite{122}

\item  numerically up to sixth order for $d=3$ in the massive scheme.~\cite
{323}
\end{enumerate}

In the two cases, only the renormalization functions $Z_{i}$ are considered.
This is because the theoretical interest is usually focused on the critical
exponents, the obtention of which comes from the knowledge of the $Z_{i}$'s.
For example the series expansion for the critical exponent $\eta $ are given
by: 
\begin{equation}
\eta (u)=\beta (u)\frac{{d}}{{d}u}\ln Z_{3}\left( u\right)  \label{eta}
\end{equation}
once considered at $u=u^{*}$.

However, we are not simply interested in the critical exponents but in
complete functions such as $\xi $ and $\chi $. Now, only in the case 2, the
renormalization conditions are such that $\xi $ and $\chi $ are known in
terms of the $Z_{i}$'s. This is not true in the case 1\footnote{%
This is why the amplitude functions are known only up to three loop order in
this scheme.~\cite{754}}.

We denote the renormalized parameters of the massive scheme by $g$ and $m$
(instead of $u$ and $t$). Their relations to the bare parameters are similar
to those given by Eqs. (\ref{W1}--\ref{R0c}) except that in addition to the
change $u\rightarrow g$, Eqs. (\ref{W3}--\ref{W4}) now read:

\begin{eqnarray}
g_{0} &=&m^{\epsilon }g\frac{Z_{1}\left( g\right) }{\left[ Z_{3}\left(
g\right) \right] ^{2}}  \label{W3nic} \\
r_{0} &=&\delta m^{2}+\frac{m^{2}}{Z_{3}\left( g\right) }  \label{dm}
\end{eqnarray}

The mass shift $\delta m^{2}$ is defined by a subtraction condition\footnote{%
Which eliminates the quadratic ultra-violet divergences occuring at $d=4$.}
which avoids the explicit consideration of $r_{0{c}}$ via Eq. (\ref{R0c}),
namely: 
\begin{equation}
\Gamma ^{(0,2)}\left( 0;m,g\right) =m^{2}  \label{m}
\end{equation}

The other subtraction conditions\footnote{%
Which eliminate the logarithmic ultra-violet divergences occuring at $d=4$.}
which define the $Z_{i}$'s read: 
\begin{eqnarray*}
\left. \frac{{d}}{{d}p^{2}}\Gamma ^{(0,2)}\left( p;m,g\right) \right| _{p=0}
&=&1 \\
\Gamma ^{(0,4)}\left( \left\{ 0\right\} ;m,g\right) &=&m^{\epsilon }g \\
\Gamma ^{(1,2)}\left( \left\{ 0,0\right\} ;m,g\right) &=&1
\end{eqnarray*}
so that the physical (bare) quantities $\xi $ and $\chi $ are given by: 
\begin{eqnarray*}
\xi ^{-1}(g) &=&m=g_{0}\frac{\left[ Z_{3}\left( g\right) \right] ^{2}}{%
gZ_{1}\left( g\right) } \\
\chi ^{-1}(g) &=&Z_{3}^{-1}m^{2}=g_{0}^{2}\frac{\left[ Z_{3}\left( g\right)
\right] ^{3}}{\left[ gZ_{1}\left( g\right) \right] ^{2}}
\end{eqnarray*}

The re-summations of the perturbative series for those quantities have been
done using the technique initiated by Le Guillou and Zinn-Justin~\cite{283}
after having accounted for the singularities of the $Z_{i}$'s at the fixed
point $g^{*}$. They may be easily treated by writing, e.g. for $Z_{3}\left(
g\right) $ which has a singularity at $g^{*}$ of the form $\left(
g^{*}-g\right) ^{\eta /\omega }$: 
\[
Z_{3}\left( g\right) =Z_{3}\left( y\right) \exp \left\{ \int_{y}^{g}\frac{%
\eta \left( x\right) }{\beta \left( x\right) }{d}x\right\} 
\]
in which $y$ is some small value of $g$, the definitions of $\beta \left(
x\right) $ and $\eta \left( x\right) $ being unchanged in their forms
compared to Eqs. (\ref{beta}, \ref{eta}). Let us mention that some
difficulties could be encountered in the resummation procedure due to
nonanalytic confluent singularities~\cite{2800} in the $\beta $-function at $%
g^{*}$, but to date they have not been numerically observed.

Thus, in the homogeneous phase, the physical quantities $\xi $ and $\chi $
can be easily estimated as functions of $g$ in the range $\left]
0,g^{*}\right[ $ from the calculated series.~\cite{323} However our aim was
to obtain those quantities as function of $\tau $ (i.e. of $r_{0}-r_{0{c}}$%
). Now, the massive framework uses $m\propto \tau ^{\nu }$ instead of $%
t\propto \tau $, therefore the linear relation to $\tau $ is lost. In order
to reintroduce it, we use the fact that, at zero-external-momenta:

\[
\left[ Z_{2}\left( g\right) \right] ^{-1}=\Gamma _{0}^{(1,2)}\left(
r_{0},g_{0}\right) =\left. \frac{\partial }{\partial r_{0}}\Gamma
_{0}^{(0,2)}\left( r_{0},g_{0}\right) \right| _{g_{0}} 
\]

Using Eqs (\ref{W3nic}--\ref{m}), we reexpress this under the following
form: 
\[
\frac{{d}\left( r_{0}/g_{0}^{2}\right) }{{d}g}=Z_{2}\left( g\right) \frac{{d}%
}{{d}g}\left\{ \frac{\left[ Z_{3}\left( g\right) \right] ^{3}}{\left[
gZ_{1}\left( g\right) \right] ^{2}}\right\} 
\]
which, after integration allows us to (implicitly) define an effective (and
nonperturbative) critical value $r_{0{c}}^{\prime }$ by reference to the
fixed point value $g^{*}$: 
\begin{equation}
\tilde{t}\left( g\right) \equiv \frac{r_{0}-r_{0{c}}^{\prime }}{g_{0}^{2}}%
=-\int_{g}^{g^{*}}{d}x\;Z_{2}\left( x\right) \frac{{d}}{{d}x}\left\{ \frac{%
\left[ Z_{3}\left( x\right) \right] ^{3}}{\left[ xZ_{1}\left( x\right)
\right] ^{2}}\right\}  \label{tnew}
\end{equation}

The integrand of Eq. (\ref{tnew}) may be estimated using the same procedure
as before and the integration has been done numerically yielding the
numerical evolution of $\tilde{t}\left( g\right) $ in the range $\left]
0,g^{*}\right[ $. The final results (the functions $\xi (\tilde{t})$ and $%
\chi (\tilde{t})$) were obtained after a fitting procedure of the implicit
form $\xi (g)$, $\chi (g)$ and $\tilde{t}\left( g\right) $. This summarizes
the calculations done in the homogeneous phase~\cite{733} which included
also the specific heat $C(\tilde{t})$ the perturbative series of which were
previously~\cite{80} extracted from the Guelph report;~\cite{323} the
calculations were performed for the symmetries $n=1,2$ and $3$.

\section{Calculations in the inhomogeneous phase and the critical bare mass}

``...{\em it is more difficult to calculate physical quantities in the
ordered phase because the theory is parametrized in terms of the disordered
phase correlation length }$m^{-1}${\em \ which is singular at }$T_{{c}}${\em %
. Also the normalization of correlation functions is singular at }$T_{{c}}$%
.''~\cite{2599}

We have calculated~\cite{736} the perturbative series for the free energy
directly at $d=3$ using the numerous already-estimated Feynman integrals of
the massive scheme~\cite{323} and new kinds of integrals which have been
estimated for the occasion.

Because the free energy is generally written in terms of $T-T_{{c}}$, we
have been led to explicitly consider the delicate question of the critical
bare mass. Indeed, it is known that the perturbative series of
super-renormalizable massless field theories (such as $\phi _{d<4}^{4}$)
develop infrared singularities which are usually simply ignored within the $%
\epsilon $-expansion framework. In 1973, using a dimensional regularization,
Symanzik~\cite{392} has shown that the critical bare mass --- which has the
form $r_{0{c}}=g_{0}^{2/\epsilon }f\left( \epsilon \right) $ in which $%
f\left( \epsilon \right) $ has poles at $\epsilon =2/k$ ($k=1,2,...\infty $)
--- is in fact an infrared regulator for the theory. However the final
result (free of infrared divergences) is no longer perturbative (e.g.,
logarithms of $g_{0}$ appear at $d=3$). Though the nonperturbative nature of 
$r_{0{c}}$ is an important aspect of the RG theory,~\cite{32} this question
may be circumvented when looking at the critical behavior. This is because $%
T_{{c}} $ is a nonuniversal quantity, thus its explicit determination is not
required, only the difference $T-T_{{c}}$ is needed. Hence, provided that
Eq. (\ref{R0c}) be again satisfied, one may redefine $r_{0{c}}$ [as it has
been done in Eq. (\ref{tnew})]. Consequently, it is allowed~\cite{736} to
perform a particular mass-shift $r_{0}=r_{0}^{\prime }+\delta r_{0}\left(
\epsilon \right) $ in such a way that $\delta r_{0}\left( \epsilon \right) $
subtracts the poles occuring at $\epsilon =1$, and to fix afterwards the
critical temperature in terms of $r_{0}^{\prime }$.

The series for the free energy have then been obtained graph by graph up to
five loops according to the following rules:

\begin{enumerate}
\item  graphs involving only $\phi ^{4}$-vertices which were already
estimated~\cite{323} with the mass-shift parameter $\delta m^{2}$ [defined
by Eq. (\ref{dm}, \ref{m})] have been re-expressed to account for the
mass-shift parameter $\delta r_{0}\left( \epsilon \right) $.

\item  New Feynman integrals at $d=3$ with their weights involving:

\begin{itemize}
\item  exclusively $\phi ^{3}$-vertices have been calculated and compared to
existing estimates.~\cite{362}

\item  $\phi ^{3}$-vertices mixed with a single $\phi ^{4}$-vertex have been
estimated for the first time for the occasion.
\end{itemize}
\end{enumerate}

Those series for the free energy have been used by Guida and Zinn-Justin~%
\cite{3639} to give an accurate estimation of the scaled equation of state.
But this kind of consideration does not account for any
correction-to-scaling terms and the comparison with experiments is not easy.
Instead, we were again interested in actually measurable quantities like the
susceptibility $\chi $, the specific heat $C$ and the spontaneous
magnetization $M$ in the inhomogeneous phase. We have not considered the
correlation length $\xi $ in this phase because the required Feynman
integrals were not calculated by Nickel et al;~\cite{323} this quantity has
been considered afterwards at $d=3 $ but up to 3-loop order only.~\cite{3556}

Because the renormalization procedure is unchanged in going into the
broken-symmetry phase, the critical singularities at the fixed point $g^{*}$
may be taken into account with the same renormalization functions $%
Z_{i}\left( g\right) $ as defined previously. But the relation between the
linear measure of the distance to $T_{{c}}$ and the unchanged renormalized
coupling $g$ is different. Instead of Eq. (\ref{tnew}) we have obtained: 
\[
\tilde{t}^{\prime }\left( g\right) =-\int_{g}^{g^{*}}{d}x\;\left\{
Z_{2}\left( x\right) \frac{{d}}{{d}x}\left\{ \frac{\left[ Z_{3}\left(
x\right) \right] ^{3}}{\left[ xZ_{1}\left( x\right) \right] ^{2}}\right\}
\left[ 1-U\left( x\right) \right] \right\} 
\]
in which $U\left( g\right) $ is given~\cite{736} as power series in $g$.

Obviously, our nonasymptotic study of the critical behavior accounts for all
the universal properties which are expected when $\tau \rightarrow 0$.
Consequently as a by-product, estimates of universal combinations of leading
critical amplitudes were given for the first time from the five loop order
at $d=3$. The recent careful re-estimations~\cite{3639,4211} of those
universal combinations from the same series have mainly reduced the
error-bars. We also gave for the first time accurate estimates of some~\cite
{34} of the universal ratios of the first confluent correction-to-scaling~%
\cite{736}.

\end{document}